\documentclass[12pt]{article} 
\usepackage{epsf}

\begin{document}
%\mbox{}
\begin{center}
{\bf \Large An Investigation of Time in Relativity}
\end{center}
\vspace*{1cm}
\begin{center}
{\bf Sidi Cherkawi Benzahra\footnote[1]{sbenzahra@wittenberg.edu or benzahra@physics.umn.edu}}
\end{center}
\begin{center}
{\large Wittenberg University}
\end{center}
\begin{center}
{\large Springfield, Ohio 45501}
\end{center}
\vspace*{1cm}
\begin{abstract}
I added an imaginary term by hand to the proper time, $\Delta T$, of special theory of 
relativity and wanted to see what 
will happen to it if I tried to get the time dilation relationship between $\Delta T^{\prime}$ 
and $\Delta T$.  I found that this added 
imaginary term of time does not have to be zero.

\end{abstract} 
\vspace*{1cm}  
\noindent
If $ \Delta T$ is the real, proper time of travel of a light beam from 
one point to the other,
\begin{equation}
\Delta t =\Delta T +i \Delta \xi
\label{firsteq}
\end{equation}  
$\Delta t$ will be the time I choose and work with in order to see what will happen to 
the imaginary part $\Delta \xi$ once I apply Einstein's special relativity. $i\Delta \xi$ here 
is an imaginary term which I added by hand.  Consider two observers $O$ and $O^{\prime}$ [1]. $O$ fires a beam of light at a mirror a 
distance L away and measures the time interval 2 $\Delta T$ for the beam to be 
reflected from the mirror and to return to $O$.  Observer $O^{\prime}$ is 
moving at a constant velocity $u$.  As seen from the point of view of $O$, the
beam is sent and received from the same point, while $O^{\prime}$ moves off
in a perpendicular direction.  The beam is sent from some point $A$ and 
received at some point $B$ a time 2 $\Delta T^{\prime}$ later, according to
$O^{\prime}$.  The distance AB is just $2u \Delta T^{\prime}$.  According to
$O$, the light beam travels a distance $2L$ in a time $2\Delta T$.  According
to $O^{\prime}$, the light beam travels a distance of $2 \sqrt{L^{2}+(u \Delta
T^{\prime})^{2}}$ in a time $2\Delta T^{\prime}$.  According to Galilean 
relativity, $\Delta T = \Delta T^{\prime}$, and $O$ measures a speed $c$ while 
$O^{\prime}$ measures a speed $\sqrt{c^{2} + u^{2}}$.  According to Einstein's 
second postulate, this is not possible--both $O$ and $O^{\prime}$ must measure 
the speed $c$.  Therefore $\Delta T$ and $\Delta T^{\prime}$ must be different.  
I can find a relationship between $\Delta T$ and $\Delta T^{\prime}$ by setting 
the two speeds equal to $c$.  According to $O$, $c=2 L/2 \Delta T$, so $L=c \Delta T$.  
According to $O^{\prime}$, $c=2\sqrt{L^{2}+ (u\Delta T^{\prime})^{2}}/2 \Delta T^{\prime}$, 
so $c\Delta T^{\prime}= \sqrt{L^{2}+(u \Delta T^{\prime})^{2}}$.  Combining these, I find
\begin{equation}
c \Delta T^{\prime}=  \sqrt{(c \Delta T)^{2}+(u \Delta T^{\prime})^{2}}
\end{equation}    
and, solving for $\Delta T^{\prime}$,
\begin{equation}
\Delta T^{\prime}= {{\Delta T }\over{\sqrt{1-u^{2}/c^{2}}}} \, . 
\end{equation}
This relationship summarizes the effect known as time dilation. But in my calculation, where
$\Delta t =\Delta T + i \Delta \xi$, I have according to $O$,
\begin{equation}
c={{L} \over {\Delta T + i \Delta \xi}}\, .
\label{landconly}
\end{equation}
I can see here in the equation just above that since I added an imaginary part to time, 
length in particular or space in general picked up an imaginary part also.  According to 
$O^{\prime}$
\begin{equation}
c={{\sqrt{L^{2}+[u(\Delta T^{\prime}+i \Delta \xi^{\prime})]^{2}}}
\over {\Delta T^{\prime} + i \Delta \xi^{\prime}}}
\label{landcandxi}\,.
\end{equation}
Taking the expression of $L$ from equation (\ref{landconly}) and inserting it in equation 
(\ref{landcandxi}) and squaring, I get
\begin{equation}
c^{2}(\Delta T^{\prime}+i \Delta \xi^{\prime})^{2}=c^{2}(\Delta T+i \Delta \xi)^{2}+
u^{2}(\Delta T^{\prime}+i \Delta \xi^{\prime})^{2} \, .
\label{realandimag}
\end{equation}
Separating the real terms from the imaginary terms of equation (\ref{realandimag}) I get

\[ [\Delta {T^{\prime}}^{2}(1-{u^{2}}/{c^{2}}) -\Delta T^{2}]-
[\Delta {\xi^{\prime}}^{2}(1-{u^{2}}/{c^{2}}) -\Delta \xi^{2}]+
2i[\Delta T^{\prime} \Delta \xi^{\prime}(1-{u^{2}}/{c^{2}})-\Delta T \Delta \xi]=0\]
I can see that time dilation relationship is given by setting the first part of the equation 
above equal to zero
\begin{equation}
\Delta T^{\prime}={{\Delta T}\over {\sqrt{1-u^{2}/c^{2}}}} \, .
\end{equation}
I can also see that there is an effect of time dilation in the imaginary part, $\Delta \xi$, 
given by 
\begin{equation}
\Delta \xi^{\prime}={{\Delta \xi}\over {\sqrt{1-u^{2}/c^{2}}}} \, .
\label{xiprime}
\end{equation}
Since a complex number vanishes only if its real and imaginary parts vanish too, I get
\begin{equation}
\Delta T^{\prime} \Delta \xi^{\prime}(1-{u^{2}}/{c^{2}})-\Delta T \Delta \xi=0 \, .
\label{imaginary}
\end{equation}
Taking the right-hand side of equation (\ref{xiprime}) and inserting it in equation 
(\ref{imaginary}), I get
\begin{equation}
[ \Delta T^{\prime} \sqrt{1-{u^{2}}/{c^{2}}} - \Delta T] \Delta \xi =0 \, .
\label{lasteq}
\end{equation}
Equation (\ref{lasteq}) is the product of two terms. Since Einstein's relativity is considered
to be true, the first part vanishes, and leaves us with the condition that $\Delta \xi$ can be
different from zero. And if $\Delta \xi$ is different from zero that makes $\Delta t$ imaginary
since $\Delta t$ is equal to $\Delta T + i \Delta \xi$ as indicated in equation (\ref{firsteq}).  
I have mathematically found that relativity allows time to have an imaginary part.   

\begin{center}
{\bf \large References}
\end{center}
[1] Kenneth Krane, Modern Physics (John Wiley and Sons, 1983).

\end{document}